\documentclass[letter,draftclsnofoot,onecolumn,12pt,oneside]{IEEEtran}
\usepackage{cite}
\usepackage{epsfig,graphics,mathrsfs,amssymb,amsmath,bm,color,yfonts,txfonts}
\usepackage{subfigure}

\begin{document}

\title{Coordinated Linear Precoding in Downlink Multicell MU-MISO OFDMA Networks }
\author{Mirza Golam Kibria, Hidekazu Murata and Susumu Yoshida\\
[2mm]
Graduate School of Informatics, Kyoto University, Kyoto, Japan\\
Email: contact-h25e@hanase.kuee.kyoto-u.ac.jp}

\maketitle

\begin{abstract}
This paper considers coordinated linear precoding in downlink multicell multiuser orthogonal frequency-division multiple access (OFDMA) network. A less-complex, fast and provably convergent algorithm that maximizes the weighted sum-rate with per base station (BS) transmit power constraint is formulated. We approximate the nonconvex weighted sum-rate maximization (WSRM) problem with a solvable convex form by means of sequential parametric convex approximation (SPCA) approach. The second order cone program (SOCP) formulations of the objective function and constraints  of the optimization problem are derived through proper change of variables, first order linear approximation and hyperbolic constraints transformation, etc. The algorithm converges to the suboptimal solution taking fewer number of iterations in comparison to other known iterative WSRM algorithms. Finally, numerical results are presented to justify the effectiveness and superiority of the proposed algorithm.
\end{abstract}

\IEEEpeerreviewmaketitle

\section{Introduction}
The weighted sum-rate maximization (WSRM) problem is known to be nonconvex and NP-hard \cite{Venturino,Wang}, even for single antenna users. Though the beamforming design methods presented in \cite{Joshi, Liu} achieve optimum capacity, these methods may be practically inapplicable since the complexity evolves exponentially with the optimization problem size. Therefore, computationally inexpensive suboptimal beamforming design is very appealing. Beamforming design based on achieving necessary conditions of optimality has been studied thoroughly in \cite{Venturino,Wang}. Importantly, the authors of \cite{Joshi} numerically prove that  the performances of the suboptimal beamforming techniques that achieve the necessary optimality conditions are indeed very close to optimal beamforming design. 

In \cite{Venturino}, the authors proposed iterative coordinated beamforming design based on Karush-Kuhn-Tucker (KKT) optimality conditions, which is not provably convergent. Alternating maximization (AM) algorithm for WSRM optimization problem is proposed in \cite{Wang}, which is based on alternating updation between a closed-form posterior conditional probability and the beamforming vectors. In \cite{Christensen,Sun,Shi}, the authors establish a relationship between weighted sum-rate and weighted minimum mean-square error (WMMSE), and solve the  WSRM optimization problem based on alternating optimization. Discrete power control based WSRM has been proposed in \cite{Zhang1}. However, all these iterative WSRM optimization designs exhibit relatively slower convergence rate in comparison to our proposed design.

In this paper, we formulate and propose a WSRM optimization solution with faster convergence for multicell orthogonal frequency division multiple access (OFDMA) system. This iterative design manipulates the sequential parametric convex approximation (SPCA) technique explored in \cite{Beck}. The SPCA based WSRM algorithm approaches the local optimal solution within a few iterations, iteratively approximating the nonconvex problem with solvable convex structure. At each step of this iterative process, the nonconvex problem is approximated with a solvable convex form and updating the acting variables until convergence. With appropriate change of variables, introducing additional optimization variable, making use of  first-order linear approximation and hyperbolic constraints transformations, we iteratively approximate the WSRM optimization problem as a second order cone program (SOCP)\cite{Lobo}.

The reminder of the paper is organized as follows. The multicell multiuser OFDMA network model and WSRM optimization framework are presented in Section II. Section III explains the process of sequential convex approximation of the nonconvex optimization problem. In Section IV, we discuss the simulation parameters and numerical results found in this work. Section V concludes the paper.\\
\textit{Notations:} $(\cdot)^{\rm{H}}$/$(\cdot)^{\rm{T}}$ stand for Hermitian-transpose/transpose operation. Gaussian distributions of real/complex random variables with mean $\mu$ and variance $\sigma^2$ is defined as $\mathcal{RN}(\mu,\sigma^2)$/$\mathcal{CN}(\mu,\sigma^2)$. Boldface lower-case/upper-case letter defines a vector/ matrix. Operator $\mathrm{vec}(\cdot)$ stacks all the elements of the argument into a column vector and $\mathrm{diag}(\cdot)$ puts the diagonal elements of a matrix in a column vector. $\mathbb{R}$ and $\mathbb{C}$ define real and complex spaces, respectively. $|\cdot|$ and $||\cdot||_2$ refer to absolute value and $l_2$ norm of the arguments, respectively. 

\section{Problem Formulation}
\subsection{System Model}
We consider an interference-limited cellular system of $M$ cells with $K$ users per cell. OFDMA multiplexing scheme with $N$ subcarriers over a fixed bandwidth is employed, while the subcarrier assignments among users within each cell are non-overlapping. Therefore, there is no intra-cell interference, only inter-cell interference is experienced by the users. The coordinated base stations (BSs) are equipped with $N_\mathrm{t}$ antennas and they are interconnected via high-capacity backhaul links. The non-cooperative users have single antenna each. Coordinated linear multiuser downlink precoding is employed at each BS. The assignment function $f(m,n)$ determines the downlink user scheduling. The assignment of user $k$ from $m$th BS on the $n$th subcarrier is defined as $k=f(m, n)$. The received data of user $k$ from cell $m$ on the $n$th subcarrier is given by
\begin{equation}
{{y}_{kmn}}={{\bm{h}}_{kmn}}{{\bm{g}}_{kmn}}{{d}_{kmn}}+\sum\limits_{\begin{smallmatrix}
 {m}'\in \mathcal{S}\backslash m \\
 {k}'=f(m', n)
\end{smallmatrix}}{{{\bm{h}}_{km'n}}{{\bm{g}}_{k'{m}'n}}{{d}_{{k}'{m}'n}}}+{{z}_{kmn}}
\end{equation}
where $k=f(m,n)$ and $\mathcal{S}$ is the set of all BSs. $\text{ }y_{kmn}\in \mathbb{C}$ denotes the received symbol for user $k$. $\bm{h}_{kmn}\in \mathbb{C}^{{1\times N_\mathrm{t}}} $ is the complex channel vector between BS $m$ and user $k$. The beamformer formed by BS $m$ to transmit data on subcarrier $n$ is denoted by $\bm{g}_{kmn}\in \mathbb{C}^{{N_\mathrm{t}}\times 1} $. $z_{kmn}\sim{\mathcal{CN}(0,1)}$ is the additive white Gaussian noise (AWGN) at user $k$. $d_{kmn}\sim{\mathcal{CN}(0,1)}$ denotes the transmitted symbol from BS $m$ to user $k$ on subcarrier $n$.
\subsection{Transmit Precoding Problem}
This paper emphasizes the linear beamformer design for sum-rate optimization in multicell multiuser OFDMA network. The design objective is to maximize the weighted sum-rate under per BS power constraints. The signal-to-interference-plus-noise ratio (SINR) of the $k$th user from cell $m$ scheduled on subcarrier $n$ is given by
\begin{equation}
\label{gamma}
{{\gamma }_{kmn}}=\frac{{{\bm{h}}_{kmn}}{{\bm{g}}_{kmn}}\bm{g}_{kmn}^{\mathrm{H}}\bm{h}_{kmn}^{\mathrm{H}}}{1+\sum\limits_{\begin{smallmatrix}
 {m}'\in \mathcal{S}\backslash m \\
  {k}'=f(m', n)
 \end{smallmatrix}}{{{\bm{h}}_{km'n}}{{\bm{g}}_{k'{m}'n}}\bm{g}_{k'{m}'n}^{\mathrm{H}}\bm{h}_{k{m}'n}^{\mathrm{H}}}}.
\end{equation}
The instantaneous downlink rate achieved by the $k$th user from cell $m$ on subcarrier $n$ is ${{c}_{kmn}}={{\log }_{2}}(1+{{\gamma }_{kmn}})$, and the instantaneous rate for user $k$ over all the subcarriers is ${{R}_{km}}=\sum\nolimits_{n\in {{\mathcal{S}}_{km}}}{{{c}_{kmn}}}$, where the summation is over all the subcarriers assigned to user $k$ from cell $m$, i.e., $n\in {\mathcal{S}_{km}}$, where ${{\mathcal{S}}_{km}}=\left\{ n|k=f(m,n) \right\}$. Let $w_{km}$ be the weight of user $k$ in cell $m$. The weight corresponding to a particular user may reflect the quality of the service it requests or its priority in the system. Then the WSRM problem is defined as
\begin{equation}
\label{optzm}
\begin{aligned}
& \underset{\mathcal{G}}{\mathop{\mathrm{maximize} }}\,\sum\limits_{m=1}^M\sum\limits_{n=1}^N{{{w}_{km}}{{c}_{kmn}}}\\
& \text{subject to} \sum\limits_{\begin{smallmatrix}
 n=1 \end{smallmatrix}}^{N}{||{{\bm{g}}_{kmn}}|{{|}_2^{2}}\le {{P}_{m, \max }}},\text{  } m=1,...,M
\end{aligned}
\end{equation}
where $\mathcal{G}:=\left\{{\bm{g}}_{kmn};\text{ }m\in M,\text{ } n\in N\right\}$ is the set of all beam forming vectors and $P_{m, \max}$ is the transmit power constraint of cell $m$. Let $\mathcal{G}_m$ be the set of beamformers for cell $m$. Since the optimization problem in \eqref{optz} is nonconvex, finding the global optimal solution is difficult and complex enough. Therefore, we focus on local optimal solution in this paper. 
\subsection{Review of  Second Order Cone Programming}
Recently, substantial progress and development have been achieved for solving a large class of optimization problems. In order to apply these algorithms, one needs to reformulate the problem into the standard form that the algorithms are capable of dealing with. Conic programs, i.e., linear programs \cite{Lobo} with generalized inequalities, are subjected to special attention. One such standard conic program is SOCP, which is of the form
\begin{equation}
\mathrm{SOCP}:\left\{ \begin{matrix}
   \hspace{1mm}\underset{\bm{x}}{\mathrm{minimize }}\ \hspace{3mm}\mathrm{Real }({{ {\bm{a}}}^{H}} {\bm{x}}) &  \\
   \hspace{-17mm}\mathrm{subject}\hspace{1mm} \mathrm{to} & \hspace{-18mm}\left[ \begin{matrix}
   {\bm{c}}_{i}^{H} {\bm{x}}+{{d}_{i}}  \\
    {\bm{D}}_{i}^{H} {\bm{x}}+{{ {\bm{b}}}_{i}}  \\
\end{matrix} \right]{{\succeq }_{\mathcal{M}}}\hspace{1mm}0,\hspace{2mm}i=1,...,U  \\
\end{matrix} \right.
\end{equation}
where $\bm{x}$ is the vector of optimization variables and $\bm{D}_i$, $\bm{b}_i$, $\bm{c}_i$, $\bm{a}_i$ are parameters with appropriate sizes. The notation $\succeq_{\mathcal{M}}$ defines the generalized inequalities:
\begin{equation}
\label{SOCP1}
\left[ (v  \hspace{3mm} {\bm{s}})^{\mathrm{T}}\right]{{\succeq }_{\mathcal{M}}}\hspace{1mm}0\Leftrightarrow || {\bm{s}}|{{|}_{2}}\le v.
\end{equation}
Hyperbolic constraints play an important role in the SOCP formulation of WSRM objective function and constraints. The hyperbolic constraints ${{\bm{w}}^{2}}\le xy,\text{~~}x\ge 0,\text{~~}y\ge 0$ with $\bm{w}\in\mathbb{R}^{1\times e}$, $x,\hspace{1mm}y\in\mathbb{R}$ and the equivalent SOCP is given by \cite{Lobo}
\begin{equation}
\label{SOCP}
{{\bm{w}}^{\mathrm{T}}}{\bm{w}}\le xy,\text{~~}{x}\ge 0,\text{~~}y\ge 0\text{~~}\Leftrightarrow \left\| \begin{matrix}
   2\bm{w}  \\
   x-y  \\
\end{matrix} \right\|_2\le x+y.
\end{equation}
\section{Sequential Parametric Convex Approximation for WSRM Problem}
As a step toward transforming the nonconvex WSRM optimization problem to SOCP\footnote{SOCP constraints are convex and can be solved using convex optimization tools such as SeDuMi \cite{Strum}. Also note that the non-overlapping subcarrier allocation in each cell does not restrict the applicability of the proposed algorithm for multiple active users in one subcarrier in one cell.} form, we reformulate the problem \eqref{optzm} into a standard form that SOCP programming is capable of dealing with. We rewrite \eqref{optzm} as 
\begin{equation}
\label{optz}
\begin{aligned}
& \underset{\bm{\mathcal{G}}}{\mathop{\mathrm{maximize} }}\,\sum\limits_{m=1}^{M}{\sum\limits_{\begin{smallmatrix}
 n=1 \\
 k=f(m, n)
\end{smallmatrix}}^{N}{{{\delta }_{kmn}}}{{\log }_{2}}(1+{{\gamma }_{kmn}})}\\
& \text{subject to} \sum\limits_{\begin{smallmatrix}
 n=1 \end{smallmatrix}}^{N}{||{{\bm{g}}_{kmn}}|{{|}_2^{2}}\le {{P}_{m,\max }}},\text{  } m=1,...,M
\end{aligned}
\end{equation}
where ${\delta }_{kmn}=w_{km},\forall n$. Let $L:=\left\{kmn,\forall m,n\text{  }|\text{  }k=f(m,n)\right\}$ and $T=MN$. Therefore, the objective function becomes a function of $T$ variables and can be expressed as
\begin{equation}
\label{optz1}
 \underset{\bm{\mathcal{G}}}{\mathop{\max }}\,\sum\limits_{t=1}^{T}{{{\delta }_{L_t}}{{\log }_{2}}(1+{{\gamma }_{L_t}})}= \underset{\mathcal{G}}{\mathop{\max }}\,\prod\limits_{t=1}^{T}{{{(1+{{\gamma }_{L_t}})}^{{{\delta }_{L_t}}}}}
\end{equation}
where $L_t$ is the $t$th set in $L$. Setting ${r}_{L_t}={{{(1+{{\gamma }_{L_t}})}^{{{\delta }_{L_t}}}}}$, we get
\begin{equation}
\label{optz2}
\begin{aligned}
&  \underset{\bm{\mathcal{G}},{{r}_{L_t}}}{\mathop{\mathrm{maximize} }}\,\prod\limits_{t=1}^{T}{{{r}_{L_t}}}\\
& \text{subject to}\text{~} C1:\sum\limits_{\begin{smallmatrix}
 n=1  \end{smallmatrix}}^{N}{||{{\bm{g}}_{kmn}}|{{|}_2^{2}}\le {{P}_{m,\max }}},\text{  } m=1,...,M\\
&\text{~~~~~~~~~~~~}C2:r_{L_t}^{{{q}_{L_t}}}\le {{\gamma }_{L_t}}+1,\text{~~}\forall L_t\in L, \text{~}t=1,...,T
\end{aligned}
\end{equation}
where $q_{L_t}=1/{\delta _{L_t}}$ and the constraints in C2 are active at the optimum. Per BS transmit power constraint C1 of \eqref{optz2} can be reformulated using $\mathrm{vec}(\cdot)$ as $|| \mathrm{vec}\left( {{\mathcal{G}}_{m}} \right)|{{|}_{2}}\le \sqrt{{{P}_{m,\max }}}$, for which the equivalent SOC according to \eqref{SOCP1} is expressed as
\begin{equation}
\label{eq44}
\left[ \begin{matrix}
 \sqrt{{{P}_{m,\max }}}    \\ 
 \mathrm{vec}\left( {{\bm{\mathcal{G}}}_{m}} \right)  \\
\end{matrix} \right]{{\succeq }_{\mathcal{M}}}\hspace{1mm}0.
\end{equation}
Further, introducing slack variables $\zeta_{L_t}\ge 0$, we can reformulate \eqref{optz2} using \eqref{gamma} as given below
\begin{equation}
\label{optz3}
\begin{aligned} 
&  \underset{\bm{\mathcal{G}}, {{r}_{L_t}}, \zeta_{L_t}}{\mathop{\mathrm{maximize} }}\,\prod\limits_{t=1}^{T}{{{r}_{L_t}}}\\
& \text{subject to}\text{~} C1:\left[ \begin{matrix}
  \sqrt{{{P}_{m,\max }}} \\
\mathrm{vec}\left( {{\bm{\mathcal{G}}}_{m}} \right) \\
\end{matrix} \right]{{\succeq }_{\mathcal{M}}}\hspace{1mm}0,\text{  } m=1,...,M\\
&\text{~~~~~~~~~~~~}C2:{{\zeta}_{{{L}_{t}}}}(r_{{{L}_{t}}}^{{{q}_{{{L}_{t}}}}}-1)^{1/2}\le {{\bm{h}}_{L_t}}{{\bm{g}}_{L_t}}\\
&\text{~~~~~~~~~~~~}C3\footnotemark:\operatorname{\mathcal{I}}\{{{\bm{h}}_{L_t}}{{\bm{g}}_{L_t}}\}=0,\text{ } \mathcal{I}\left\{x\right\}=\text{Imaginary part of }x\\
&\text{~~~~~~~~~~~~}C4:\sqrt{1+\sum\limits_{\begin{smallmatrix}{m}'\in \mathcal{S}\backslash m \\
  {k}'=f(m', n)
 \end{smallmatrix}}{{{\bm{h}}_{km'n}}{{\bm{g}}_{k'm'n}}\bm{g}_{k'm'n}^{\mathrm{H}}\bm{h}_{km'n}^{\mathrm{H}}}}\le {{\zeta }_{{{L}_{t}}}}
\end{aligned}
\end{equation}
\footnotetext{For any $\phi$, we have $|\bm{h}_{L_t}\bm{g}_{L_t}|^2=|\bm{h}_{L_t}\bm{g}_{L_t}e^{j\phi}|^2$. Therefore, by choosing $\phi$ such that $\mathcal{I}\{{{\bm{h}}_{L_t}}{{\bm{g}}_{L_t}}\}=0$ does not affect the optimality of \eqref{optz3}. } 
Let $\bm{H}_{\operatorname{int}}\in\mathbb{C}^{{(M-1)\times N_\mathrm{t}}} $ and $\bm{G}_{\operatorname{int}}\in\mathbb{C}^{N_\mathrm{t} \times(M-1)}$ be the collected channel and beamforming matrices, respectively, containing the channels from all interfering BSs and beamforming vectors corresponding to the constraint C4 of \eqref{optz3}. Therefore, we can write the constraint C4 as ${{\left\| {{\left[ \begin{matrix}1 & \mathrm{diag}\left( {{\bm{H}}_{\operatorname{int}}}{{\bm{G}}_{\operatorname{int}}} \right)  \\ \end{matrix} \right]}^{\mathrm{T}}} \right\|}_{2}}\le {{\zeta }_{{{L}_{t}}}}$, which is equivalent to the SOCP constraint
\begin{equation}
\label{equat33}
\left[ \begin{matrix}
   {{\zeta }_{{{L}_{t}}}}  \\
   {{\left[ \begin{matrix}
   1 & \mathrm{diag}\left( {{\bm{H}}_{\operatorname{int}}}{{\bm{G}}_{\operatorname{int}}} \right)  \\
\end{matrix} \right]}^{\mathrm{T}}}  \\
\end{matrix} \right]{{\succeq }_{\mathcal{M}}}\hspace{1mm}0.
\end{equation}
Constraints C1, C3-C4 of \eqref{optz3} are convex, hence require no approximation. However, C2 is still nonconvex. To make use of the SPCA technique to approximate C2 as a convex constraint, we break C2 of \eqref{optz3} and reformulate as 
\begin{equation}
\label{eq5}
v_{{{L}_{t}}}^{1/2}{{\zeta }_{{{L}_{t}}}}\le {{\bm{h}}_{L_t}}{{\bm{g}}_{L_t}},\hspace{1mm}\forall {{L}_{t}}\in L
\end{equation}
\begin{equation}
\label{eq6}
r_{{{L}_{t}}}^{{{q}_{{{L}_{t}}}}}\le {{v}_{{{L}_{t}}}}+1.
\end{equation}
Though both \eqref{eq5} and \eqref{eq6} are still nonconvex, yet this formulation facilitates us to use the established convex approximation methods. First, we consider the convex approximation of \eqref{eq5}. Defining $\mathcal{Q}({{\zeta}_{{{L}_{t}}}},{{v}_{{{L}_{t}}}})=v_{{{L}_{t}}}^{1/2}{{\zeta }_{{{L}_{t}}}}$ with $v_{{{L}_{t}}},\text{ }{{\zeta }_{{{L}_{t}}}}\ge0$, we approximate $\mathcal{Q}({{\zeta}_{{{L}_{t}}}},{{v}_{{{L}_{t}}}})$ with its convex upper estimate function \cite{Beck} $G({{\zeta }_{{{L}_{t}}}},{{v}_{{{L}_{t}}}},{{\theta }_{{{L}_{t}}}})$ as
\begin{equation}
\label{eq1}
G({{\zeta }_{{{L}_{t}}}},{{v}_{{{L}_{t}}}},{{\theta }_{{{L}_{t}}}})\triangleq \frac{1}{2}\left( \frac{{{v}_{{{L}_{t}}}}}{{{\theta }_{{{L}_{t}}}}}+{{\theta }_{{{L}_{t}}}}\zeta _{{{L}_{t}}}^{2} \right).
\end{equation}
Hence, $\mathcal{Q}({{\zeta }_{{{L}_{t}}}},{{v}_{{{L}_{t}}}})\le G({{\zeta }_{{{L}_{t}}}},{{v}_{{{L}_{t}}}},{{\theta }_{{{L}_{t}}}})$, $\forall {\theta }_{{L}_{t}}\ge 0$. At the optimum, $\mathcal{Q}({{\zeta}_{{{L}_{t}}}},{{v}_{{{L}_{t}}}})=G({{\zeta}_{{{L}_{t}}}},{{v}_{{{L}_{t}}}},{{\theta}_{{{L}_{t}}}})$  when ${{\theta }_{{{L}_{t}}}}=\sqrt{{{v}_{{{L}_{t}}}}}/{{\zeta }_{{{L}_{t}}}}$. This point can be reached in an iterative way by intuitively updating the variables until we obtain the KKT points of  \eqref{optz3}. Convex overestimation of $\mathcal{Q}({{\zeta}_{{{L}_{t}}}},{{v}_{{{L}_{t}}}})$ allows us to express equation \eqref{eq5} as hyperbolic constraints, and the SOCP representation for the corresponding hyperbolic constraints is given by
\begin{equation}
\label{equat2}
{{\left\| {{\left[ {{\zeta }_{{{L}_{t}}}}\sqrt{\frac{\theta _{{{L}_{t}}}}{2}}\text{~~}({{\bm{h}}_{L_t}}{{\bm{g}}_{L_t}}-\frac{{{v}_{{{L}_{t}}}}}{2\theta _{{{L}_{t}}}}-1)] \right]}^{\mathrm{T}}} \right\|}_{2}}\le ({{\bm{h}}_{L_t}}{{\bm{g}}_{L_t}}-\frac{{{v}_{{{L}_{t}}}}}{2\theta _{{{L}_{t}}}}+1)
\end{equation}
which can be equivalently expressed as SOCP constraint as
\begin{equation}
\label{equat22}
\left[ \begin{matrix}
   {{\bm{h}}_{{{L}_{t}}}}{{\bm{g}}_{{{L}_{t}}}}-\frac{{{v}_{{{L}_{t}}}}}{2\theta _{{{L}_{t}}}}+1  \\
   {{\left[ {{\zeta }_{{{L}_{t}}}}\sqrt{\frac{\theta _{{{L}_{t}}}}{2}}\hspace{4mm}({{\bm{h}}_{{{L}_{t}}}}{{\bm{g}}_{{{L}_{t}}}}-\frac{{{v}_{{{L}_{t}}}}}{2\theta _{{{L}_{t}}}}-1) \right]}^{\mathrm{T}}}  \\
\end{matrix} \right]{{\succeq }_{\mathcal{M}}}\hspace{1mm}0.
\end{equation}
Now, let us turn our focus on \eqref{eq6} and we notice that the term $r_{{{L}_{t}}}^{{{q}_{{{L}_{t}}}}}$ is a differentiable function. To arrive at SOCP, we scale all $q_{L_t}$ such that $q_{L_t}<1$ so as to make the function $r_{{{L}_{t}}}^{{{q}_{{{L}_{t}}}}}$ concave.  For a differentiable function $\mathcal{V}$ with $(\forall x,y\in\text{domain }(\mathcal{V}))$, the first order condition for concavity says that a function $\mathcal{V}$ is concave if and only if the gradient line is the global over-estimator of the function \cite{Lobo}. The function $\mathcal{V}(x)+{{\nabla}_{x}}\mathcal{V}{{(x)}^{T}}(y-x)$ is defined as the first order approximation to the function at $x$, where ${{\left({{\nabla}_{x}}\mathcal{V}(x)\right)}_{i}}=\frac{\partial \mathcal{V}(x)}{\partial {{x}_{i}}}$. Correspondingly, we approximate $r_{k}^{{{q}_{k}}}$ with its concave over-estimator as follows
\begin{equation}
\label{equat1}
\begin{aligned}
& r_{{{L}_{t}}}^{{{q}_{{{L}_{t}}}}}-r_{{{L}_{t}},i}^{{{q}_{{{L}_{t}}}}}\le {{q}_{{{L}_{t}}}}r_{{{L}_{t}},i}^{{{q}_{{{L}_{t}}}}-1}({{r}_{{{L}_{t}}}}-{{r}_{{{L}_{t}},i}})\\
& \text{i.e., } {{v}_{{{L}_{t}}}}\ge {{q}_{{{L}_{t}}}}r_{{{L}_{t}},i}^{{{q}_{{{L}_{t}}}}-1}({{r}_{{{L}_{t}}}}-{{r}_{{{L}_{t}},i}})+r_{{{L}_{t}},i}^{{{q}_{{{L}_{t}}}}}-1 \text{ }(\text{using \eqref{eq6}})\\
\end{aligned}
\end{equation}
and iteratively solve until convergence in parallel with \eqref{equat2}. In fact, it is the linearization of $r_{{{L}_{t}}}^{{{q}_{{{L}_{t}}}}}$ around the point ${{r}_{{{L}_{t}},i}}$, where ${{r}_{{{L}_{t}},i}}$ is the value of ${{r}_{{{L}_{t}}}}$ at the $i$th iteration. Both \eqref{equat2} and \eqref{equat1} are increasing function; however, they are upper bounded by the per BS power constraints.
Now, we turn our attention to the objective function. There are two possible ways to convexify the objective function of \eqref{optz3} and the methods are as follows

\textbf{Method 1:} The geometric mean (GM) of the optimization 
variables $\chi=({{r}_{L_1}}{{r}_{L_2}}...{{r}_{L_{T}}})^{1/{T}}$ is concave when ${{r}_{L_t}}\succeq 0,\forall L_t$. Maximizing the GM of the optimization variables will serve the same weighted sum-rate as maximizing the product of the optimization variables as long as the variables are nonnegative affine\cite{Lobo}, hence we can rewrite the objective function as
\begin{equation}
\label{eqGM}
\underset{\bm{\mathcal{G}},{{r}_{L_t}},\zeta_{L_t}}{\mathop{\mathrm{maximize} }}\,\prod\limits_{t=1}^{T}{{{r}_{L_t}}}:\Leftrightarrow \underset{\bm{\mathcal{G}},{{r}_{L_t}},\zeta_{L_t}}{\mathop{\mathrm{maximize} }}\,\prod\limits_{t=1}^{T}({{r}_{L_t}})^{1/T}.
\end{equation}
Using the CVX\cite{CVX} solver with SeduMi, a disciplined convex programming, we can directly use the GM of the optimization variables as objective function. We refer this method as SPCA-GM and it is not in SOCP form.

\textbf{Method 2:} The second approach is based on transforming
the product of the optimization variables into hyperbolic constraints, which also admit SOCP representation. Thus, we require to reformulate the problem by introducing new variables and by incorporating hyperbolic constraints.
Let us define the set of new variables as $\psi$. During the transformation process the variables are assigned values at $\log_2{T}$ stages. For simplified analysis, let $T=2^{p}$, where $p$ is a real positive quantity. The transformation procedure is provided below.
\[ \begin{aligned} \hline
& \textbf{Procedure 1:} \text{ for hyperbolic constraints transformation} \\ \hline
& {\textbf{Initialize:  }} \psi _{t}^{p}={{r}_{{{L}_{t}}}},\text{  }t=1,...,T\text{  and  } p={\log}_{2}(T)\\
& \textbf{for  }j=p,p-1,...,1 \\
& {{\left( \psi _{i}^{j-1} \right)}^{2}}\le \psi _{2i-1}^{j}\psi _{2i}^{j},\text{  }i=1,...,{{2}^{j-1}}\\
& \textbf{end}\\
 \hline
\end{aligned} \] 
The overall SPCA-WSRM algorithm is summarized here:
\[ \begin{aligned} \hline
& \textbf{SPCA-WSRM algorithm:}\\ \hline
& 1.\text{ }{\textbf{Initialize:  }} I_{max}, (\theta_{L_t}^{i}, r_{L_t}^{i}, {\zeta }_{{{L}_{t}}}^i), i=0\\
& 2.\text{ }\textbf{repeat}\\
& 3.\text{ }\textbf{solve the following:}\\
& \text{~  }\underset{\bm{\mathcal{G}},{{r}_{L_t}},\zeta_{L_t}}{\mathop{\mathrm{maximize} }}\,\chi \text{ (if GM approach\textbf{ (Method 1)} is used) or}\\
& \text{~  } \underset{\bm{\mathcal{G}},{{r}_{L_t}},\zeta_{L_t}, v_{L_t}, \psi_{L_t}}{\mathop{\mathrm{maximize} }}\,\psi^0 \text{ (if SOCP approach\textbf{ (Method 2)} is used)}\\
& \text{~  }\text{subject to}\\
& \text{~  }\text{}C1:{\textbf{Procedure 1}} \text{with \eqref{SOCP} (ignore if \textbf{Method 1} is used)}.\\
& \text{~  }\text{}C2:\left[ \begin{matrix}
  \sqrt{{{P}_{m,\max }}} \\
\mathrm{vec}\left( {{\bm{\mathcal{G}}}_{m}} \right) \\
\end{matrix} \right]{{\succeq }_{\mathcal{M}}}\hspace{1mm}0,\text{  } m=1,...,M\\
& \text{~  }\text{}C3:\left[ \begin{matrix}
   {{\bm{h}}_{{{L}_{t}}}}{{\bm{g}}_{{{L}_{t}}}}-\frac{{{v}_{{{L}_{t}}}}}{2\theta _{{{L}_{t}}}^{i}}+1  \\
   {{\left[ {{\zeta }_{{{L}_{t}}}}\sqrt{\frac{\theta _{{{L}_{t}}}^{i}}{2}}\hspace{4mm}({{\bm{h}}_{{{L}_{t}}}}{{\bm{g}}_{{{L}_{t}}}}-\frac{{{v}_{{{L}_{t}}}}}{2\theta _{{{L}_{t}}}^{i}}-1) \right]}^{\mathrm{T}}}  \\
\end{matrix} \right]{{\succeq }_{\mathcal{M}}}\hspace{1mm}0\\
& \text{~  }\text{}C4:\operatorname{\mathcal{I}}\{{{\bm{h}}_{L_t}}{{\bm{g}}_{L_t}}\}=0\\
& \text{~  }\text{}C5: {{v}_{{{L}_{t}}}}\ge {{q}_{{{L}_{t}}}}r_{{{L}_{t}},i}^{{{q}_{{{L}_{t}}}}-1}({{r}_{{{L}_{t}}}}-{{r}_{{{L}_{t}},i}})+r_{{{L}_{t}},i}^{{{q}_{{{L}_{t}}}}}-1\\
& \text{~  }\text{}C6:\left[ \begin{matrix}
   {{\zeta }_{{{L}_{t}}}}  \\
   {{\left[ \begin{matrix}
   1 & \mathrm{diag}\left( {{\bm{H}}_{\operatorname{int}}}{{\bm{G}}_{\operatorname{int}}} \right)  \\
\end{matrix} \right]}^{\mathrm{T}}}  \\
\end{matrix} \right]{{\succeq }_{\mathcal{M}}}\hspace{1mm}0\\
& \text{~  }\text{}C7:\psi_{L_t}\ge0, r_{L_t}\ge0 \text{ implicit constraints }\\
& 4.\text{ }\text{denote }(r_{L_t}^{i+1}, \zeta_{L_t}^{i+1}, v_{L_t}^{i+1})= \text{optimal values at step 3.}\\
& 5.\text{ }\theta _{{L_{t}}}^{i+1}=\sqrt{v_{L_t}^{i+1}}/\zeta_{L_t}^{i+1},\text{ }\text{ }i=i+1\\
& 6.\text{ }\textbf{until }\text{convergence or } i=I_{max}\\
\hline
\end{aligned} \] 
The objective function emerges to be a one variable function defined as $\psi_1^0=\psi^0$, which is obtained at the final stage of hyperbolic constraint formulation described in \textbf{Procedure 1}. Finally, applying \eqref{SOCP} yields the SOCP formulations for $2^p-1$ hyperbolic equations of \textbf{Method 2}. 
It is worth noting that this algorithm is inspired by \cite{Beck, Chris, Tran} and is similar to \cite{Tran}, which proposes the SPCA based algorithm for multicell MU-MISO networks. However, we formulate and propose the SPCA based algorithm with GM approach for multicell OFDMA networks and resolve two practical limiting factors related to the algorithm implementation, which are not addressed in \cite{Tran} to make the algorithm more general, especially when the problem size is comparatively larger.

The initial ${{\theta }_{{{L}_{t}}}}$s are very crucial to the feasibility and convergence of the SPCA-WSRM algorithm. It could be possible that for some cases, the randomly generated ${{\theta }_{{{L}_{t}}}}$s can lead to infeasible solution at the first iteration. To make sure that the
algorithm is feasible at the first step, we follow the steps in \textbf{Procedure 2}  to find good initial ${{\theta }_{{{L}_{t}}}}$s.

The other numerical issue that is not addressed in \cite{Tran} is the situation when one or some of the
$v_{L_t}$s become zero, i.e., no power on that or those particular subcarriers of the corresponding cell. It is usual that some of the subcarriers may not get any power due to limited BS power if we recall the mechanism of water-filling algorithm. However, when such situation arises, we have noticed numeral instability. We encounter the problem of dividing by zero since we need to calculate $1/\theta_{L_t}$. In order to avoid this situation 
we slightly modify the imposed constraints on $v_{L_t}$ such as $v_{L_t}\ge \varepsilon$ (e.g., $\varepsilon$=0.0001) so that we bypass the numerical problem.
By this constraint, the algorithm yields a solution that is close to the original one without encountering the numerical instability.

\[ \begin{aligned} \hline
& \textbf{Procedure 2:} \text{ Proposal for generating initial values of  } {{\theta }_{{{L}_{t}}}}\\ \hline
& {\textbf{Step 1: }}\text{Generate channel-matched beamforming vectors so} \\
& \text{that per BS power constraint is satisfied for all cells, i.e.,} \\
& {{\bm{g}}_{kmn}}=\sqrt{{{{P}_{m,\max }}}/{N}}({{{\bm{h}}_{kmn}}}/{||{{\bm{h}}_{kmn}}||_2}), \forall m,n \text{ and }k=f(m,n) \\
& {\textbf{Step 2: }}\text{ Use C4 of \eqref{optz3} to find }{\zeta }_{{L}_{t}}^0\text{ by replacing inequality } \\
& \text{with equality. } \\
& {\textbf{Step 3: }}\text{Calculate }r_{L_{t}}^0\text{ from  C2 of \eqref{optz3} putting the absolute }\\
& \text{value of } {{\bm{h}}_{L_t}}{{\bm{g}}_{L_t}}. \\
& {\textbf{Step 4: }}\text{Find }v_{L_t}\text{ using \eqref{eq5}}.\text{ Finally the initial value of }\theta_{L_t} \\
& \text{is obtained as }\theta_{L_t}^0=\sqrt{{{v}_{{{L}_{t}}}}}/{{\zeta }_{{{L}_{t}}}^0}.\\
 \hline
\end{aligned} \] 
\section{Numerical Results}
The performance of the proposed algorithm is analyzed on a cellular network with 3 coordinated BSs and 2 users per cell, with 1-cell frequency reuse factor, via Monte-Carlo simulations. The distance between adjacent BSs is 1000 m. The users are uniformly distributed around its own BS within a circular annulus of external and internal radii of 1000 m and 500 m, respectively. Like the paper \cite{Venturino}, frequency-selective channel coefficients over 64 subcarriers are modeled as  
\begin{equation}
{{\bm{h}}_{kmn}}={{\left(200\frac{1}{{{l}_{km}}}\right)}^{3.5}}{{\Phi }_{kmn}}{{\Lambda }_{kmn}}
\end{equation}
where ${l}_{km}$ is the distance between BS $m$ and user $k$. $10{{\log }_{10}}({{\Phi }_{kmn}})$ is distributed as $\mathcal{RN}(0,8)$, accounting for  log-normal shadowing and ${{\Lambda }_{kmn}}\sim\mathcal{CN}(0,1)$ accounts for Rayleigh fading. All the BSs are subjected to the  equal maximum power constraint, i.e., ${{P}_{m,\max }}={{P}_{\max }},\forall m$. We also consider that perfect channel state information (CSI) is available both at the BSs and users. The initial-user assignment is performed randomly. We consider $N_\mathrm{t}=2$ and use CVX\cite{CVX} package for specifying and solving convex programs.
 \begin{figure}
  \centering
   \includegraphics[scale=0.15]{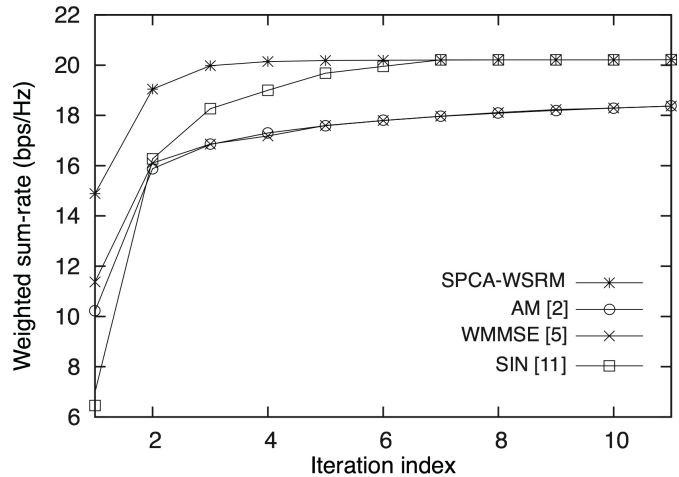}
   \caption{Convergence rate comparison for different WSRM algorithms.  }
   \label{BER1}
 \end{figure}
 \begin{figure}
  \centering
   \includegraphics[scale=0.15]{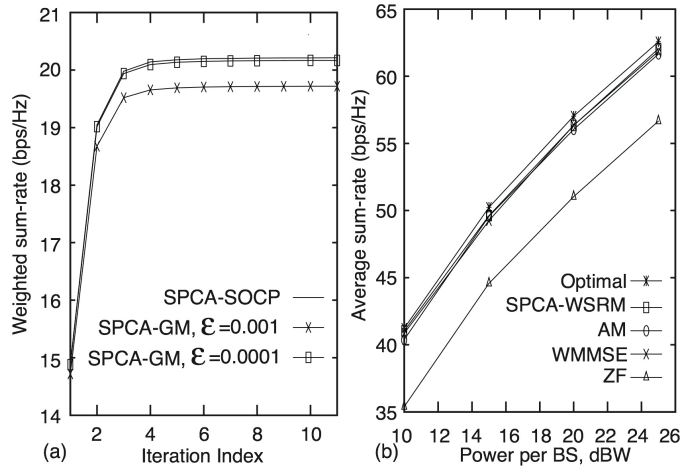}
   \caption{(a). Performance comparison between \textbf{Method 1} and \textbf{Method 2}, (b). Average sum-rate performances for different WSRM algorithms.  }
   \label{BER2}
 \end{figure}
 
In Fig.~\ref{BER1}, we compare the WSR achieved by all schemes as a function of the number of iterations required to acquire steady output for a random channel realization. The maximum power limit for all the BS is set to 20 dBW, i.e., ${{P}_{\max }}=20\hspace{1mm}\mathrm{dBW} $. It is easily noticed that SPCA-WSRM algorithm converges within few iterations, while the AM and WMMSE are still far away from convergence level of SPCA-WSRM. This phenomenon may be attributed to the fact that AM-WSRM optimization requires alternation between a closed-form posterior conditional probability update and updating the beamforming vectors, while the WMMSE algorithm relies on the relationship between mutual information and minimum mean-square error (MMSE), and alternates between updating of transmit and receive beamformers. As a result, comparatively slower convergences are observed. However,  good initial values for the variables involved in WMMSE accelerate the convergence rate. Though SIN algorithm, which is also based on convex approximation of the precoder covariance matrices, has similar convergence performance to SPCA-WSRM. However, the per iteration running time is much higher. 

Fig.~\ref{BER2}a compares the WSR performances for the two different methods described in the previous section. For both methods, we generate the initial values of ${{\theta }_{{{L}_{t}}}}$s using \textbf{Procedure 2} and modify the constraints on $v_{L_t}$s as we discussed. Although both methods exhibit same WSR performance for higher values of $\varepsilon$, the per iteration running time for \textbf{Method 1} is much longer than \textbf{Method 2}. This is attributed to the fact that the solver internally transforms the GM to hyperbolic constraints in each iteration. We have observed that the algorithm provides feasible solution to the optimization problem all the times. It is obvious that the larger the value of $\varepsilon$, the bigger performance gap between \textbf{Method 2} and \textbf{Method 2} evolves.

Finally, in Fig.~\ref{BER2}b, we compare the average sum-rate ($w_{km}$=1) performances for various precoding strategies as a function of per BS transmit power. The suboptimal solutions achieved by SPCA-WSRM algorithm and other techniques such as AM and WMMSE are indeed very close to the optimal precoding performance obtained from \cite{Joshi}. However, AM and WMMSE require a large number of iterations to reach their respective suboptimal levels.
\section{Conclusions}
In this paper, we study the WSRM optimization problem for a multicell OFDMA multiplexing system. We formulate and propose an SPCA based convex approximation of the optimization problem, which is known to be nonconvex and NP-hard. This iterative SOCP optimization is provably convergent to the local optimal solution. Some numerical issues related to the algorithm implementation are also discussed. Particularly, in terms of convergence rate, this algorithm exhibits excellent performance and outperforms some previously analyzed solutions to the WSRM optimization problem.
\ifCLASSOPTIONcaptionsoff
  \newpage
\fi

\end{document}